\documentclass[journal]{IEEEtran}
\usepackage{amsmath,amsfonts}
\usepackage{algorithmic}
\usepackage{algorithm}
\usepackage{array}
\usepackage[caption=false,font=normalsize,labelfont=sf,textfont=sf]{subfig}
\usepackage{textcomp}
\usepackage{stfloats}
\usepackage{url}
\usepackage{verbatim}
\usepackage{graphicx}
\usepackage{cite}
\usepackage{acronym}
\usepackage[hidelinks]{hyperref}
\usepackage{multirow}
\usepackage{booktabs}
\newacro{TTS}{text-to-speech}
\newacro{CV}{computer vision}
\newacro{minDCF}{minimum detection cost function}
\newacro{EER}{equal error rate}
\newacro{SRE}{NIST speaker recognition evaluation}

\newacro{ASR}{automatic speech recognition}
\newacro{DINO}{DIstillation with NO labels}
\newacro{LDA}{linear discriminant analysis}
\newacro{PLDA}{probabilistic \ac{LDA}}
\newacro{SSL}{Self-supervised learning}
\newacro{AAM}{additive angular margin}
\newacro{MoCo}{momentum contrast}

\newacro{LR}{logistic regression}
\newacro{SVM}{support vector machine}
\newacro{PLDA}{probabilistic linear discriminant analysis}
\newacro{SV}{speaker verification}
\newacro{SER}{speaker emotion recognition}
\newacro{PCA}{principal component analysis}
\newacro{CV}{cross validation}
\newacro{AD}{Alzheimer's disease}
\newacro{MMSE}{mini-mental status evaluation}

\begin{document}

\title{Non-Contrastive Self-supervised Learning for Utterance-Level Information Extraction from Speech}

\author{Jaejin Cho,~\IEEEmembership{Student Member, IEEE}, Jes\'us Villalba,~\IEEEmembership{Member, IEEE}, Laureano Moro-Velazquez,~\IEEEmembership{Member, IEEE}, Najim Dehak,~\IEEEmembership{Senior Member, IEEE}
        
\thanks{All authors are associated with the Department of Electrical and Computer
Engineering, and the Center for Language and Speech Processing (CLSP),
Johns Hopkins University, Baltimore, MD, 21218 USA while Jes\'us Villalba and Najim Dehak are also affiliated to the Human
Language Technology Center of Excellence, Johns Hopkins University,
Baltimore, MD, 21218, USA}
\thanks{This project was supported by NSF Award 1816165.}}

\markboth{IEEE JSTSP 2022 Early Access Version}%
{Shell \MakeLowercase{\textit{et al.}}: A Sample Article Using IEEEtran.cls for IEEE Journals}


\maketitle

\begin{abstract}
In recent studies, self-supervised pre-trained models tend to outperform supervised pre-trained models in transfer learning. In particular, self-supervised learning of utterance-level speech representation can be used in speech applications that require discriminative representation of consistent attributes within an utterance: speaker, language, emotion, and age. Existing frame-level self-supervised speech representation, e.g., wav2vec, can be used as utterance-level representation with pooling, but the models are usually large. There are also self-supervised learning techniques to learn utterance-level representation. One of the most successful is a contrastive method, which requires negative sampling: selecting alternative samples to contrast with the current sample (anchor). However, this does not ensure that all the negative samples belong to classes different from the anchor class without labels. This paper applies a non-contrastive self-supervised method to learn utterance-level embeddings. We adapted DIstillation with NO labels (DINO) from computer vision to speech. Unlike contrastive methods, DINO does not require negative sampling. We compared DINO to x-vector trained in a supervised manner. When transferred to down-stream tasks (speaker verification, speech emotion recognition, and Alzheimer’s disease detection), DINO outperformed x-vector. We studied the influence of several aspects during transfer learning such as dividing the fine-tuning process into steps, chunk lengths, or augmentation. During fine-tuning, tuning the last affine layers first and then the whole network surpassed fine-tuning all at once. Using shorter chunk lengths, although they generate more diverse inputs, did not necessarily improve performance, implying speech segments at least with a specific length are required for better performance per application. Augmentation was helpful in speech emotion recognition.
\end{abstract}


\section{Introduction}
As larger scale computation becomes possible, systems based on deep learning started to outperform the systems with classical machine learning algorithms in many applications. In deep learning, however, training a well-performing network from scratch usually requires a lot of data with labels. Getting labels for large data requires a lot of time and cost, accompanying human annotators most of the times. Thus, much larger amount of data are left without labels. To adapt to this situation, an increasing number of studies have started to propose new techniques to exploit unlabeled data during the last few years. \ac{SSL} is one of the techniques, which learns from the structure of the data itself, not relying on labels.
In \cite{devlin-etal-2019-bert,lee2020biobert,NEURIPS2020wav2vec2,chen2020simclrv2,NEURIPS2020BYOL,caron2021dino}, fine-tuned/post-processed \ac{SSL} models outperformed supervised methods when the same amount of labeled data is used.

In the speech processing field, we can divide \ac{SSL} largely into two groups regarding scale of interest to extract representation: frame-level and utterance-level \ac{SSL}. The frame-level \ac{SSL} techniques aim to learn representation in an unsupervised way mainly to solve sequence prediction problems such as \ac{ASR} and phoneme recognition~\cite{ravanelli2020paseplus,chung2019apc,liu2020mockingjay,liu21npc,liu2021tera,NEURIPS2020wav2vec2,hsu2021hubert}. However, the learned frame-level representations also can be used as an utterance-level representation with a pooling layer, possibly with a fine-tuning process to generate better utterance-level embeddings~\cite{yang21superb}. Although this frame-level \ac{SSL} are flexible in its usages for speech applications, they are usually large in the model sizes.
The utterance-level \ac{SSL} learns a representational vector per utterance to tackle problems such as language/accent/speaker identification/verification, and emotion recognition~\cite{hsu2017unsupervised,stafylakis2019ss_spkemb,cho20tts_spkid,peng2020mixture,huh2020augmentation,xia2021moco_spkemb,zhang2021contrast_spkid}. Although the models trained with the utterance-level \ac{SSL} cannot be used in frame-level, they usually have smaller model sizes compared to the frame-level \ac{SSL} models.

In utterance-level \ac{SSL}, contrastive losses are the most popular~\cite{huh2020augmentation,xia2021moco_spkemb,zhang2021contrast_spkid}. The loss guides the model to make the current sample (anchor) closer to an augmented version of the current sample (positive sample) in the embedding space while pushing the positive sample farther from the negative ones. Here, the negative samples are those desired to be different semantically from the positive sample. Since the samples are unlabeled, most works compose negative samples by picking random samples different from the anchor. In this way of random sampling, however, we cannot be sure if all the negative samples belong to the classes different from the positive sample's class. For example, when the anchor, thus also the positive sample, is an utterance from speaker A, there is a chance that some of the randomly picked negative utterances also come from speaker A. This could adversely affect the model training since the contrastive loss pushes the positive sample and negative sample farther to each other in the embedding space.

On the other hand, non-contrastive methods, do not require negative samples, so they are free from the issue above. Moreover, non-contrastive methods have shown similar or better performance compared to contrastive methods~\cite{NEURIPS2020BYOL,caron2021dino}. Considering these, in this study we adapt a non-contrastive \ac{SSL} method originally proposed in computer vision to speech, \ac{DINO}~\cite{caron2021dino}, that outperformed the previous \ac{SSL} methods in many computer vision tasks.

Several works train supervised models to use their hidden representations in target tasks that usually suffer from data scarcity. In~\cite{tits2018frozenasr2emo}, the authors used a pre-trained ASR model as a feature extractor for emotion recognition tasks. In \cite{pappagari2020xvecmeetemo, padi2021xvec2emo,pappagari2021copypaste}, the authors utilized a pre-trained speaker classification models that trained on large data for \ac{SER}. Especially in~\cite{pappagari2020xvecmeetemo,pappagari2021copypaste}, three x-vector models were trained from scratch to distinguish different emotion classes, and evaluated on three different emotion corpora. However, the models performances were always worse than the models fine-tuned from the pre-trained x-vector models trained to distinguish speakers using a lot of speaker-labeled data. We think this is because the x-vector model has too many parameters to train only with an emotion corpus, which usually have far less samples compared to the number of samples in speaker ID corpus. Using a multitask objective, e.g., emotion classification loss plus \ac{AD} classification loss, can be one way to increase the total number of samples, but an \ac{AD} dataset as a medical data also usually has a small number of samples, thus making the multitask objective training still not solving the data scarcity problem better than transfer learning. In \cite{pappagari2020adress,pappagari21adresso}, the authors used pre-trained speaker classification, and encoder-decoder ASR models for \ac{AD} detection. 

In this paper, we compared a self-supervised and supervised pre-trained model in three speech applications, \ac{SV}, \ac{SER}, and \ac{AD} detection through extensive experiments. In the pre-training stage, we adapted a non-contrastive self-supervised learning technique, \ac{DINO}, first proposed in computer vision, to the speech domain for the self-supervised pre-trained model, and used \ac{AAM}~\cite{deng2019arcface} loss-based speaker classification model for the supervised pre-trained model. Once the pre-training is done, we transferred the two models to each speech application either by using them as a feature extractor or fine-tuning them with a newly added linear layer. The contributions of this paper are following:
\begin{itemize}
  \item We adapted \ac{DINO} to the speech domain, showing promising results throughout the experiments.
  \item We studied how the ratio of short segment length to long segment length affects the \ac{DINO} embedding learning.
  \item We studied the effect of \ac{PCA} when pre-trained models are used as feature extractors and classifiers are trained on a small amount of the data.
  \item We took 2 steps in fine-tuning a pre-trained model with a newly added affine layer for a target task: first tuning the affine layers and then the whole network together. This 2-stage fine-tuning worked better than fine-tuning the whole network at once, possibly reducing deviation of parameters from the pre-trained model, which aids to utilize knowledge from pre-training with large data.
  \item We analyzed the influence of different chunk lengths and chunk padding techniques in the results.
  \item We checked how noise data augmentation affects for \ac{SER} and \ac{AD} detection.
\end{itemize}
Through the series of experiments above, we found that \ac{DINO} enables building \ac{SV} systems without speaker labels, and given the fixed amount of the speaker-labeled data, it improves x-vector for \ac{SV} by providing a better initialized model. When transferred to \ac{SER} and \ac{AD} detection tasks, \ac{DINO} outperformed x-vector. The findings suggest that \ac{DINO} effectively learns general utterance-level embedding that includes not only speaker information but also emotion and \ac{AD}-related information without any labels.

The organization of this paper is as follows: We first introduce x-vectors, which were used as a baseline in our experiments (Section II). Then, we describe the \ac{DINO} with how it was adapted to the speech domain (Section III). We explain how to transfer the pre-trained models in Section IV. The \ac{SV} back-ends and an iterative process to build an \ac{SV} system without labels explained in Section V and VI, respectively. Section VII presents our experimental setup, which is followed by the results and analyses from three speech applications (Section VIII, IX, X). Finally, we conclude the paper in Section XI with the following discussion in Section XII.

\begin{table}\caption{LResNet34 encoder architecture. The same architecture was used in both x-vector and \ac{DINO}}
\label{tab:lresnet_enc_arch}
\centering
\begin{tabular}{|c|c|c|}
\hline
Component  & Layer   & Output Size   \\ \hline \hline
\multirow{6}{*}{\begin{tabular}[c]{@{}l@{}}Frame-level\\conv blocks \end{tabular}} & $3 \times 3, 16$                                                                          & $T \times 80$        \\ \cline{2-3} 
& \begin{tabular}[c]{@{}l@{}} $\begin{bmatrix} 3 \times 3, 16 \\ 3 \times 3, 16 \end{bmatrix} \times 3$ \end{tabular}             & $T \times 80$        \\ \cline{2-3} 
& \begin{tabular}[c]{@{}l@{}} $\begin{bmatrix} 3 \times 3, 32 \\ 3 \times 3, 32 \end{bmatrix} \times 4$, stride 2\end{tabular}   & $\frac{T}{2} \times 40$    \\ \cline{2-3} 
& \begin{tabular}[c]{@{}l@{}} $\begin{bmatrix} 3 \times 3, 64 \\ 3 \times 3, 64 \end{bmatrix} \times 6$, stride 2\end{tabular}   & $\frac{T}{4} \times 20$     \\ \cline{2-3} 
& \begin{tabular}[c]{@{}l@{}} $\begin{bmatrix} 3 \times 3, 128 \\ 3 \times 3, 128 \end{bmatrix} \times 3$, stride 2\end{tabular} & $\frac{T}{8} \times 10$     \\ \cline{2-3} \hline
Pooling   &   global mean + standard deviation  &   $2560$  \\ \hline
Embedding & fully connected & 256 \\\hline
\end{tabular}
\end{table}

\section{x-Vectors}
In this paper, we used the x-vector embedding as a baseline~\cite{snyder2017xvecarch, snyder2018xvector}. The x-vector network is trained with speaker ID labels to classify speakers. The network is composed of mainly three parts: frame-level encoder, pooling layer, and classification head. The frame-level encoder processes the given input utterance sequence, e.g., MFCC or Mel filter banks, into frame-level representations. This frame-level features are fed into a pooling layer that aggregates the features into a fix-dimensional vector. Finally, the classification head processes the fix-dimensional vector to calculate speaker class logit. The network was trained with cross entropy loss in the past but nowadays its variant, the \ac{AAM} softmax loss~\cite{deng2019arcface}, is popular, showing better performance in speaker verification task. The \ac{AAM} softmax loss aims to cluster embeddings of the same class, and separate embeddings of the different classes in a hyper-sphere with an additive angular margin. We used the \ac{AAM} softmax loss in all experiments unless otherwise mentioned.

x-vectors usually correspond to the hidden vector representations extracted from the penultimate layer in the classification head. There has been many variation proposed over the years regarding x-vector network architectures \cite{he2016resnet,vaswani2017transformer,villalba2018jhumitNISTsre2018,nj2021spktransf}. In this paper, we used a light version of ResNet34 (LResNet34) proposed in~\cite{villalba2018jhumitNISTsre2018} with small modification considering limited computation resources. Table~\ref{tab:lresnet_enc_arch} shows the configuration excluding the final layer, which we call LResNet34 encoder throughout the paper. The final layer we used for x-vector was an affine layer that projects the LResNet34 encoder output into the logit vector with its dimension equal to the number of speakers used in training.

\section{Distillation with NO labels in speech} \label{sec:dino_in_speech}
\vspace{-0.05in}
\begin{figure}[htp]
    \centering
    \vspace{0.05\textwidth}
    \includegraphics[width=1.0\linewidth]{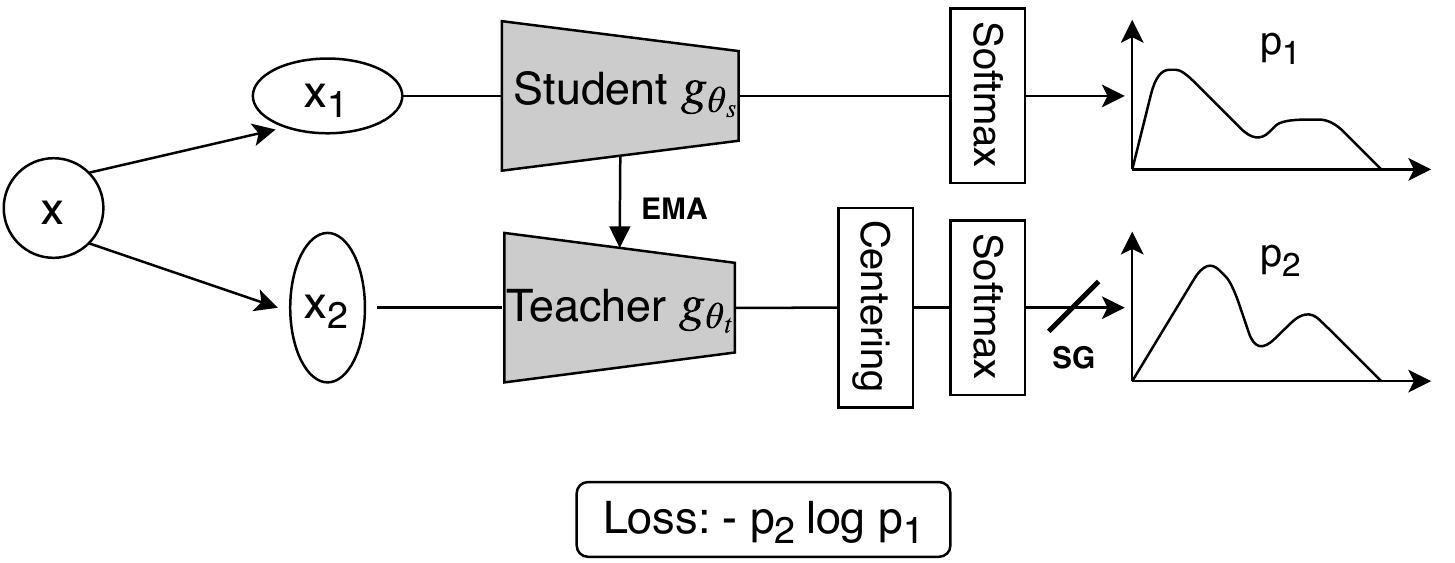}
    \vspace{0.05\textwidth}
    \includegraphics[width=1.0\linewidth]{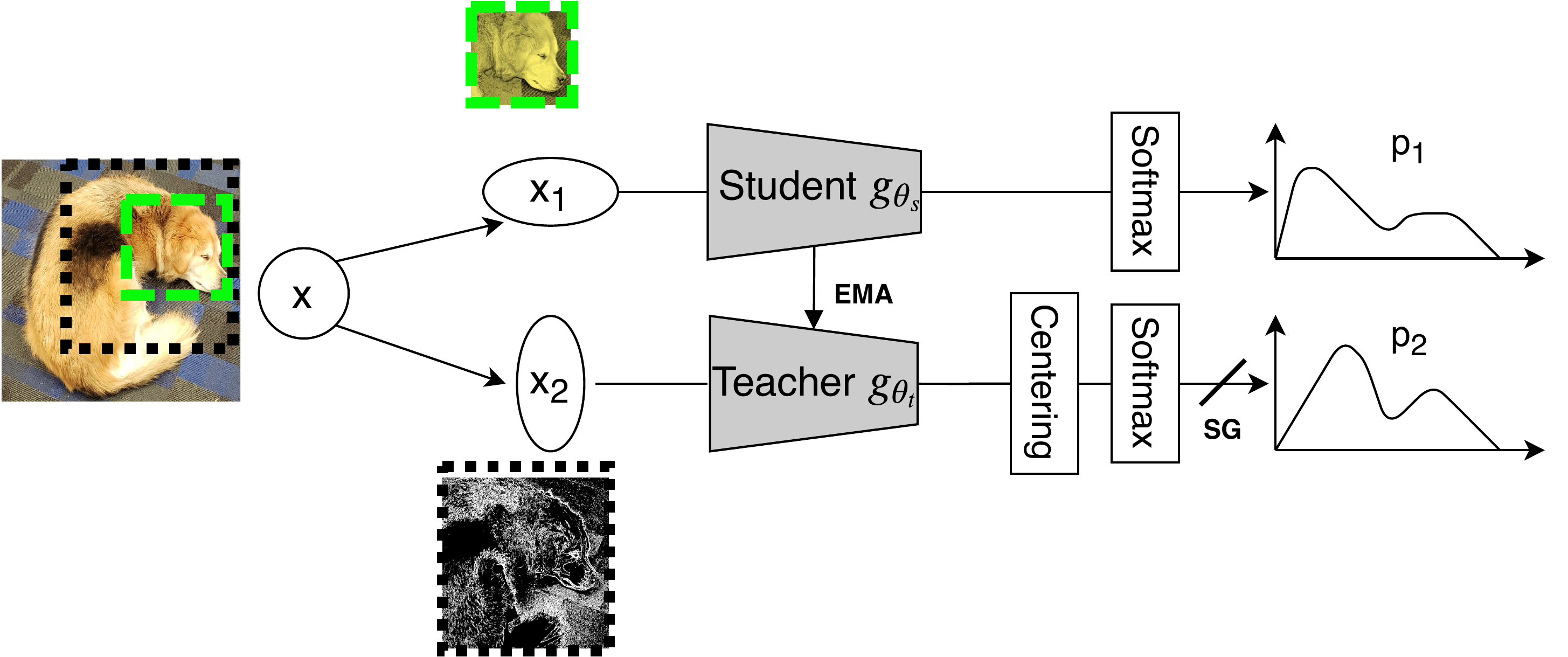}\label{subfig:dino_image}
    \includegraphics[width=1.0\linewidth]{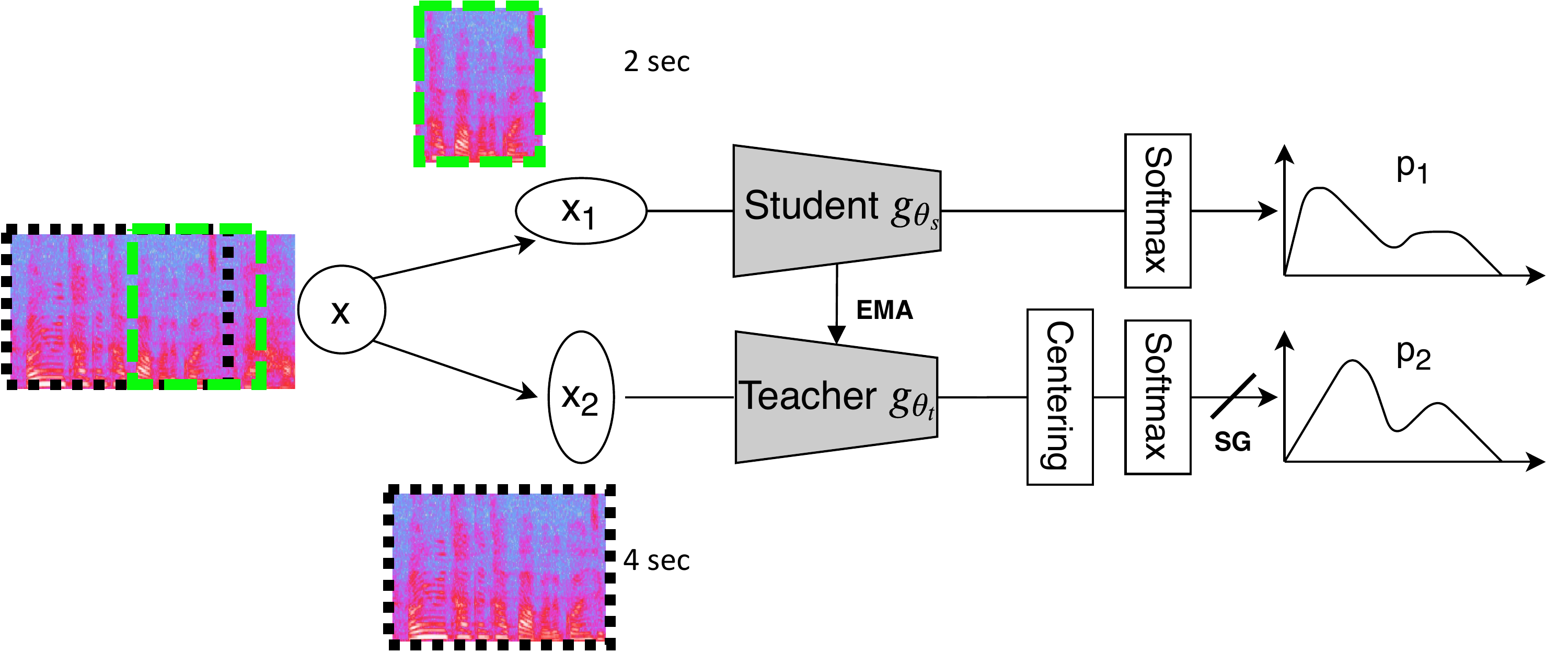}\label{subfig:dino_speech}
    \caption{\ac{DINO} diagram. EMA stands for exponential moving average. SG stands for stop gradient. The figure shows a single augmented pair ($\textrm{x}_1$,$\textrm{x}_2$)  for simplicity. 
    We call shaded parts student and teacher networks, while the upper branch including the student network is called student branch and the lower branch teacher branch.
    Second and third sub-figures show how input manipulation is different between computer vision and speech.}
    \label{fig:dino}
\end{figure}

\ac{DINO} was proposed in computer vision to maximize the similarity between feature distributions of differently augmented images from an original image~\cite{caron2021dino}. This technique assumes that augmented images from an image keep the same semantic information. For example, after cropping two different portions from a dog image and transform one to monochrome while applying jitters to the other, they are still dog images. This idea can be similarly applied to the speech domain. Assuming that one utterance only contains the speech from one speaker, different segments extracted from that utterance and transformed with noise augmentation, have the same characteristics that are maintained within the utterance, such as speaker information, accent/language, emotion, speech pathology, and age.

\ac{DINO} loss uses a knowledge distillation concept, where outputs of a teacher network are used as ground truths to optimize a student network. DINO, however, trains both networks in parallel differently from typical knowledge distillation that uses a pre-trained teacher network. Fig.~\ref{fig:dino} depicts DINO's training scheme.
First, a given utterance is augmented into a set of differently transformed segments, $S$. In detail, each segment is extracted at random position from the given utterance as either a short or long segment. Then, a noise such as babbling, music, environmental noise, or room impulse response effect is applied to each segment differently. The set $S$ includes two long augmented segments, $x^{l}_1$ and $x^{l}_2$, and multiple short augmented segments, e.g., four in our experiments. All the augmented segments go through the student branch, while only the long augmented segments go through the teacher branch. The network in each branch first encodes information in the corresponding augmented segments into the embedding vectors. Then, the following module called projection head in the network outputs a $K$ dimensional feature which will be normalized with a temperature softmax over the feature dimension as shown in Eq.~\ref{eq:tmp_sfmx_student}.
\begin{equation}
  p_1(x)^{(i)} = \frac{\exp(g_{\theta_s}(x)^{(i)} / \tau_s)}{\sum_{k=1}^K \exp(g_{\theta_s}(x)^{(k)} / \tau_s)},
\label{eq:tmp_sfmx_student}
\end{equation}  
where $i$ is an index for an element in the softmax vector, $g_{\theta_s}$ is a student network, and $\tau_s$ is student temperature parameter that controls smoothness of the softmax in the student branch. 
The same equation holds for $p_2$ with $\tau_t$ for the teacher branch.

The student network $\theta_s$ is trained by minimizing the loss,  
\begin{equation}
    \min_{\theta_s} \sum_{x \in \{x^{l}_1, x^{l}_2\}} \quad \sum_{\substack{x' \in S\\x'\neq \, x}} \quad H(p_2(x|\theta_t), p_1(x'|\theta_s)).
\label{eq:loss}
\end{equation}
where $H(a, b) = - a \cdot\log b$ is cross-entropy, and $p_1(.)$ and $p_2(.)$ are the softmax output of the student and teacher branch respectively. This loss makes the embeddings of all augmented versions of the utterance close. This is a form of Mean Teacher self-distillation~\cite{NIPS2017meanteacherbetter}, which relies on two assumptions. First, the long segments, used as teacher network inputs, will produce better representations than the short segments. Second, the teacher network is always better than the student during the training, as explained below.

The neural network architecture $g$ for student and teacher models is composed of a backbone $f$ followed by a projection head $h$: $g = h \circ f$.
The backbone can be any encoder that converts a sequence of vectors into a fixed-dimensional vector, e.g., using a pooling layer, while the projection head consists of fully connected layers with non-linear activations, which outputs a $K$ dimensional feature. In our work, we used LResNet34 encoder in Table~\ref{tab:lresnet_enc_arch} for the backbone, which was also used for x-vector.
The student and teacher networks have the same architecture, initialized with the same parameters while they are updated in different ways during training. The student network is updated by gradient descent while the teacher network is updated by an exponential moving average on the student parameters, i.e., $ \theta_t \leftarrow \lambda \theta_t + (1-\lambda) \theta_s$ where $\lambda$ is a teacher momentum hyper-parameter. Parameter averaging is known to produce a better model~\cite{polyak1992acceleration,NIPS2017meanteacherbetter}, and this is also the case with the teacher network to be better than the student network during the training. The student model aims at the distribution from the teacher network to improve.

To avoid a model to find trivial solutions, i.e., having distributions where one dimension is dominant or having uniform distributions, \textit{centering} and \textit{sharpening} are applied. \textit{centering} prevents one dimension from dominating by calculating a center, as in Eq.~\ref{eq:center_update}, and subtracting it from the logit before the softmax in the teacher network as $g_{\theta_t}(x) \leftarrow g_{\theta_t}(x) - c$ where $g_{\theta_t}$ is the teacher network. 
\begin{equation}
c \leftarrow m c + (1-m) \frac{1}{B} \sum_{i=1}^B g_{\theta_t}(x_i),
\label{eq:center_update}
\end{equation}
where $m$ is a center momentum, and $B$ is batch size.
However, \textit{centering} encourages a uniform distribution, and thus \textit{sharpening} is also applied to encourage peaky distributions. This is done by setting a low value for the temperature $\tau_t$ in the teacher softmax normalization.

\section{Transfer Learning}
Transfer learning is a type of learning that utilizes knowledge, gained from solving one problem, to other related problems. In our work, we could make use of the pre-trained models, e.g., by x-vector or \ac{DINO} training, to several speech applications such as speech emotion recognition or \ac{AD} detection. This is because the ability of the pre-trained models that summarize information in a given utterance could be useful to those applications. We explore 2 ways in this paper: 1) using the pre-trained models as utterance-level embedding extractors, and 2) fine-tuning the models with a newly added affine layer that is task-specific.

For the first method, a pre-trained model is frozen to extract utterance-level embeddings from the target domain data. Then, the embeddings are used to train a new model that can solve the problem of interest. Comparing different pre-trained models in this setup could measure which model generates better embeddings that characterize relevant information for the problem of interest. However, this method has limitation in that the pre-trained models cannot be tuned anymore.

In contrast, the fine-tuning method can also tune the parameters of the pre-trained model, expecting the embeddings to improve as well as the classifier. In fine-tuning, an additional module that is specific to the target task is added to a part of the pre-trained model. In our experiments, for example, we only kept the LResNet34 encoders in Table~\ref{tab:lresnet_enc_arch} from both x-vector and \ac{DINO} pre-trained models and added an affine layer after the encoder that projects the embedding into a logit vector with its dimension as the number of classes in the target domain. One important goal of the fine-tuning is tuning the parameters of the pre-trained model for the target task while utilizing the knowledge gained from the pre-training stage the most. In other words, if the parameters in the pre-trained model change too much during the tuning, the final model could not exploit the knowledge learned in the pre-training stage. To keep the parameters of the pre-trained model from changing too much, we take 2-steps during the fine-tuning stage by first tuning only the affine layers, i.e., the last embedding layer in the LResNet34 encoder in Table~\ref{tab:lresnet_enc_arch} and the newly added affine layer, and then tuning the whole part together. In this way, it prevents large gradients of the newly added layer from propagating back to the previous layers that causes the parameters to change largely.

\section{Speaker Verification Back-ends}
When enrollment and test utterances are given in a trial, the speaker verification system calculates the score to determine whether the pair comes from the same speaker or not, with a threshold. To do this, the front-end system, e.g., x-vector and DINO networks, extracts a fixed dimensional embedding vector per utterance from its encoder. Then, the embeddings of the enrollment and test utterances are given to a back-end to calculate the score. In this section, we explain two main back-ends based on cosine scoring and \ac{PLDA}~\cite{ioffe2006plda}.

\subsection{Cosine Scoring}
Given a pair of embeddings extracted from enrollment and test utterances, cosine scoring calculates the cosine similarity of the pair. This does not require any training to score a given pair.

\subsection{\ac{PLDA}}
\ac{PLDA} is a generative model. A speaker embedding $w_{ij}$ from the session $i$ of the speaker $j$ is written as below:
\begin{equation}
w_{ij}=\mu+\textbf{\textrm{V}}y_i+\epsilon_{ij}
\end{equation} 
where $\mu$, $\textbf{\textrm{V}}$, $y_i$, and $\epsilon_{ij}$ is a speaker-independent term, low-rank matrix of eigen-voices, speaker factor vector, and offset vector, respectively. $y_i \sim N(0,\textbf{\textrm{I}})$ and carries the speaker information. $\epsilon_{ij} \sim N(0,\textbf{\textrm{S}}_W)$, which describes the variability between different sessions of the same speaker. $\textbf{\textrm{S}}_W$ is the within-class covariance while the between-class covariance is calculated as $\textbf{\textrm{S}}_b=\textbf{\textrm{V}}\textbf{\textrm{V}}^T$

\ac{PLDA} is scored by computing the ratio between the likelihood of the trial embeddings given the target hypothesis and the corresponding likelihood given the non-target hypothesis. If the speaker in the enrollment and test embeddings is the same, both embeddings have the same speaker factor $y$ but different channel offsets. If the the enrollment and test embeddings are from different speakers, they have different speaker factors and different channel offsets. The \ac{PLDA} training requires speaker labeled data.

\section{Iterative pseudo-labeling and fine-tuning in \ac{SV}}
Iterative pseudo-labeling and fine-tuning is first proposed in~\cite{thienpondt2020idlab}. This is a unsupervised learning pipeline that trains a x-vector model that is continually updated with refined pseudo speaker labels over cycles. The process starts from a initial model trained in a self-supervised manner. In our paper, \ac{DINO} was used for the initial model. Then, the new larger model is trained based on pseudo speaker labels generated from the initial model. In detail, the labels are generated by clustering the embeddings extracted from the initial model, after which assigning pseudo speaker IDs to the clusters. Then, the trained new larger model is used to extract embeddings for another turn of clustering followed by pseudo-labeling. This process is repeated until the speaker verification performance converged on a validation data. There is a last stage called the robust training stage where a new bigger model is trained with a larger margin in the \ac{AAM} loss than the margin in the last cycle. This stage uses the pseudo labels generated from the model in the last cycle.

\section{Experimental setup}

\subsection{Datasets}

\subsubsection{Model pre-training data}
VoxCeleb2 \textit{dev} was used to pre-train two models, x-vector and \ac{DINO}.
VoxCeleb2 \textit{dev} set has 1,092,009 utterances from 5,994 speakers~\cite{Chung18voxceleb2}. It consists of conversational speech utterances with moderate noise, which were processed from interview videos of 5,994 celebrities uploaded on Youtube, covering diverse ethnicity.

\subsubsection{Transfer learning data} 
For transfer learning, we used the VoxCeleb subset, IEMOCAP~\cite{busso2008iemocap}, and ADReSSo2021~\cite{LuzEtAl21ADReSSo} for \ac{SV}, \ac{SER}, and \ac{AD} detection, respectively.

\emph{VoxCeleb subset:}
We used VoxCeleb1~\cite{nagrani2017voxceleb1}, VoxcSRC-21\footnote{https://www.robots.ox.ac.uk/~vgg/data/voxceleb/competition2021.html} $val$ and $test$ trials.
VoxCeleb1 \textit{dev} of the verification split was used for fine-tuning pre-trained models or training \ac{PLDA} back-ends while VoxCeleb1 \textit{test} was employed to validate the developed speaker verification systems. The Voxceleb1 corpus was collected similarly as Voxceleb2. VoxCeleb1 \textit{dev} has 148,642 utterances from 1,211 speakers and VoxCeleb1 \textit{test} trials were generated from 4,874 utterances from 40 speakers. VoxcSRC-21 $val$ and $test$ trials are from VoxCeleb Speaker Recognition Challenge2021(VoxSRC-21), where the challenge has a special focus on multi-lingual verification. These two trials, in addition to the VoxCeleb1 \textit{test} trials,  were used to evaluate a progressively updated model during iterative pseudo-labeling and fine-tuning.

\emph{IEMOCAP:}
In \ac{SER} transfer learning experiments, IEMOCAP was employed. IEMOCAP dataset is composed of utterances which were processed from 5 multi-modal dyadic conversational sessions. 5 female and 5 male actors participated, and in each session, one male and female actor acted for emotional conversations about pre-defined topics. Sessions were segmented into utterances manually, after which each utterance was annotated by at least 3 annotators for one of 8 emotion classes.
Conversations were either scripted or improvisational. 
In this work, we chose a subset of data consisting of 4 emotions: angry, sad, happy, and neutral. The total number of utterances was 4490. As the number utterances in this data is small, we ran 5-fold \ac{CV} where each fold is composed of one session to avoid speakers in common between folds. The selection of the subset and the fold division follows what's employed in \cite{pappagari2021copypaste}.

\emph{ADReSSo2021:}
For Alzheimer's disease (AD) detection  transfer learning experiments, we used the training subset of the dataset provided for the ADReSSo2021 challenge where the labels represent whether participants have \ac{AD} or not. The recordings employed in this study include a picture description, used for \ac{AD} detection. In general, each recording contains an interaction between a participant and an investigator with different types of background noises. The used subset has 87 recordings from speakers with \ac{AD} and 79 from control subjects. Considering the small amount of the utterances in the data, we ran 10-fold \ac{CV} as used in~\cite{pappagari21adresso} where each fold has class-balanced distribution.

\subsubsection{Utterance duration}
We provide the sample duration histograms for two corpora, IEMOCAP, and ADReSSo2021 used for \ac{SER} and \ac{AD} detection, respectively. The average duration of samples are much longer in ADReSSo2021 since one sample is one recording session in ADReSSo2021 rather than one utterance as in IEMOCAP.

\subsection{Feature Extraction}

Throughout the experiments, the speech waveform was sampled 16 kHz rate. Before feeding the samples to the models, we first removed silence portions using an energy-based voice activity detector (VAD), and extracted 80-dimensional log Mel filter bank features with 25 ms frame length at every 10 ms. Mean and variance normalization with its moving window of 150-frame length was applied to the log Mel filter bank sequence.

\begin{figure}
\centering
\includegraphics[width=0.40\textwidth]{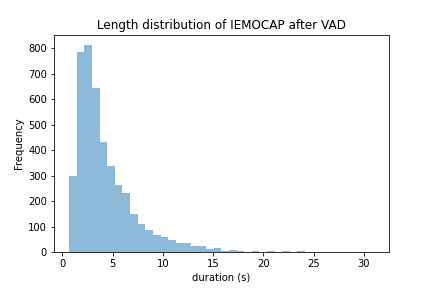}\label{subfig:IEMOCAP_dur_hist_aftervad}
\includegraphics[width=0.40\textwidth]{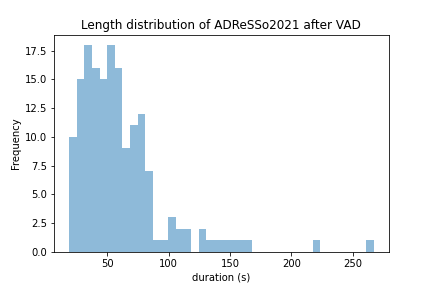}\label{subfig:ADReSSo2021_dur_hist_aftervado}
\caption{Histogram of sample duration after VAD}
\label{fig:utt_dur_hist}
\end{figure}

\subsection{Model pre-training} \label{subsec:model_pretraining}
We pre-trained a x-vector and DINO models on Voxceleb2, later to transfer them into three down-stream tasks. The LResnet34 encoder in Table~\ref{tab:lresnet_enc_arch} was used within both models. The x-vector model was trained using speaker labels while the \ac{DINO} model did not use any labels.

In x-vector, the \ac{AAM} loss~\cite{deng2019arcface} was followed after the LResnet34 encoder with feature scale $s$ and margin $\mu$ set to 30 and 0.3, respectively. The margin value warmed up linearly from 0 to 0.3 during first 20 epochs. The training ran over 70 epochs with the effective batch size as 512. We used the Adam optimizer~\cite{kingma2015adam} with its learning rate of 0.05, \textit{betas}=(0.9,0.95), \textit{weight\_decay}=1e-5, and \textit{amsgrad}=True. The learning rate warmed up for 1000 steps to 0.05 and kept the value for 40000 steps, after which it started to reduce exponentially with 0.5 decay rate, 8000 decay steps, and 1e-5 learning rate lower bound. For the training inputs, we used 4 s chunks extracted from the utterances at random positions. 

In \ac{DINO}, the LResnet34 encoder $f$ outputs 256-dimensional embedding vector given an input speech segment. The output embedding is fed into a projection head $h$, which comprised three 2048-dimensional affine layers and the following $l_\textrm{2}$ normalization and a weight normalized affine layer with 65536 output dimension. The student and teacher networks have this same architecture, initialized with the same parameters while they are updated in different manners as explained in Section~\ref{sec:dino_in_speech}. Finally, the 65536-dimensional output logit is normalized with a temperature softmax~\cite{caron2021dino}. The training ran for 70 epochs with its effective batch size 128 while the last layer was frozen for the first epoch. The Adam optimizer was used with learning rate of 0.0025, \textit{betas}=(0.9,0.95), \textit{weight\_decay}=1e-4, and \textit{amsgrad}=True. The learning rate was linearly increased for the first 10 epochs to 0.0025. After this, it decayed with a cosine schedule with the minimum learning rate as 1e-6. Student temperature $\tau_s$ and teacher temperature $\tau_t$ were set to 0.1 and 0.04, respectively. The teacher momentum $\lambda$ was cosine-scheduled from 0.996 to 1. The center momentum $m$ in \textit{centering} was set to 0.9. We extracted four 2 s and two 4 s chunks from a given utterance for short and long segments, respectively. The utterances shorter than 4 s after VAD were removed to avoid padding,  which discarded 6.4\% of the original samples.

For both models, the same data augmentation policy was applied on the fly to the extracted chunks. First, reverberation was applied with 0.45 probability using an impulse response randomly selected from small, medium, and real rooms\footnote{\url{http://www.openslr.org/  resources/28/rirs_noises.zip}}~\cite{ko2017rir}. Then, we used three noise types from the MUSAN corpus: babble, music and generic noise. A randomly selected noise type among the three was added to the whole part of a given chunk with 0.7 probability. SNR-level was contr  olled when a noise is added within 3 to 18, 3 to 18, and 0 to 18 dB for babble, music and generic noise, respectively.

\subsection{Speaker verification}
\subsubsection{Frozen}
In this experiment, we freezed the LResNet34 encoders of two pre-trained models and extracted embeddings from VoxCeleb1 \textit{test}. The embeddings were fed to two back-end systems with the trial list for verification. We compared two back-ends: \ac{PLDA} trained on VoxCeleb1 \textit{dev}, and cosine scoring that did not require any training.
\subsubsection{Fine-tuning} \label{SV_FT}
In fine-tuning, we added to the LResNet34 encoder of each pre-trained model, an affine layer as the last layer that projects the embedding into a logit where its dimension is the number of speakers in VoxCeleb1 \textit{dev}, 1211. Cross entropy or \ac{AAM} loss were used for the loss. We compared two ways of fine-tuning: 1) fine-tuning the pre-trained LResNet34 encoder with the added affine layer at once (\textit{FT1}), 2) fine-tuning only the affine layers first, and then fine-tuning the whole parts further (\textit{FT2}).

The fine-tuning ran for 50 epochs with learning rate 0.0001 and the effective batch size of 128. The number of epochs and learning rate were less than the ones used in the pre-training stage. The Adam optimizer was used with \textit{betas}=(0.9,0.95), \textit{weight\_decay}=1e-5, and \textit{amsgrad}=True. The learning rate decreased by a factor of 10 if the validation loss does not decrease after 10 epochs. In the \textit{FT2} case, this optimization policy was employed in each stage. 
When the pre-trained models were fine-tuned with \textit{FT2} using the \ac{AAM} loss ($\textit{FT2}_{\textrm{AAM}}$), the same optimization policy was used as in the x-vector pre-training except the less number of epoch and less learning rate as 50 and 0.0001, respectively and 128 batch size. The stage 2 of $\textit{FT2}_{\textrm{AAM}}$ started from the margin $\mu$ set to 30.

\subsubsection{Iterative pseudo-labeling and fine-tuning}
The training data was fixed as VoxCeleb2 \textit{dev} without speaker labels for the whole pipeline. We used the pre-trained \ac{DINO} model as our initial model. For pseudo-labeling, we used k-means clustering with 50k means, followed by agglomerative hierarchical clustering (AHC) with the number of clusters as 7500, which was heuristically determined for VoxCeleb2 \textit{dev}~\cite{thienpondt2020idlab}. The k-means clustering was used to make AHC computationally viable. Indices of the clusters were used as pseudo speaker labels for the supervised x-vector model training. We trained a new larger x-vector model, ResNet34~\cite{he2016resnet}, with the \ac{AAM}~\cite{deng2019arcface} loss based on pseudo speaker labels generated using the initial \ac{DINO} model. The newly trained model was then used for a new turn of pseudo-labeling and fine-tuning where the process repeated twice until the speaker verification performance converged. A 2-second segment was extracted per utterance to be used as a training sample.

Finally, in the robust training stage, we used a new model, even larger than ResNet34, Res2Net50~\cite{gao2019res2net} with 26 for the width of filters (in the first residual bottleneck block) and 4 for the scale. The model was trained in a supervised way with pseudo labels generated from the ResNet34 model trained over the 3 cycles above. After the first 30 epochs of training, the post-pooling layers of the model were fine-tuned with a larger margin, 0.5, in the \ac{AAM} loss. A 2-second segment was extracted from each utterance for a training sample, while 3-second segments were used in the large margin fine-tuning.

The learned embedding in each stage was evaluated for speaker verification on VoxCeleb1 \textit{test} (VoxCeleb1\_test\_o), VoxSRC-21 \textit{val}, or VoxSRC-21 \textit{test} trials.

\subsection{Speech emotion recognition and \ac{AD} detection}
\subsubsection{Frozen}
In this experiment, we freezed the two pre-trained models and extracted embeddings from the encoders. The embeddings were used to train three classifiers: \ac{LR}, \ac{SVM} with a RBF kernel, and \ac{PLDA} classifiers. We also experimented with \ac{PCA} to reduce embedding dimension before feeding it to classifiers. We did grid search for the \ac{PCA} dimension from 15 to 75 with step size of 5.
\subsubsection{Fine-tuning}
We added an affine layer as a last layer to the pre-trained LResNet34 encoder of each pre-trained model. The added layer projects the embedding into a logit with the dimension as the number of classes, 4 and 2 for \ac{SER} and \ac{AD} detection, respectively. Cross entropy was used for the loss. We compared two fine-tuning methods, \textit{FT1} and \textit{FT2}.

The number of epochs, the optimizer setting, learning rate, and learning rate scheduling were the same as used in Section~\ref{SV_FT} while the effective batch size was set to 64 and 16 for \ac{SER} and \ac{AD}, respectively. In the \textit{FT2} case, this optimization policy was applied in each stage.

In the fine-tuning experiments, we explored several hyper-parameters to see how they influence the training results. First one is how different chunk lengths affect the fine-tuning performance. This could be considered a type of data augmentation. In emotion recognition, for example, if there is a labeled utterance of 7 s, we might predict the same emotion from only a 4 s chunk of it, which can be sampled multiple times within the utterance.
Next, we compared zero-padding and repeating for short utterances when using chunks of a fixed length during training. In detail, if we set a chunk length as $T$ during training, the utterance shorter than the length $T$ need to be padded in some way for batch processing. In our experiments, we compared two methods, padding them with zeros and padding them by repeating the utterances. 
Lastly, we checked if noise data augmentation helps in fine-tuning using the same data augmentation policy explained in Section~\ref{subsec:model_pretraining}.

\section{Speaker Verification Results}
\subsection{Dependency of a ratio of long and short segment lengths}
In this experiment, we examined two aspects in \ac{DINO} embedding learning: 1) length of long augmented segment inputs (TABLE~\ref{tab:dino_longseg_len}) and 2) ratio of long and short augmented segment inputs (TABLE~\ref{tab:dino_seg_ratio}). The evaluation was done for \ac{SV} on VoxCeleb1 \textit{test} in \ac{EER}(\%).

As shown in TABLE~\ref{tab:dino_longseg_len}, when the long segment length is 2 s, it performed worse than 3 s or 4 s. This suggests that the long segment needs to be longer than some length, e.g., 2s in our case. As the long segment length goes from 3 to 4 s, the performance gain gets smaller.

To check if changing the ratio of long segment length to short segment length helps when the long segment length is short, e.g., 2s, we also ran (2 s, 1 s) whose result is shown in the last row in TABLE~\ref{tab:dino_longseg_len}. However, it performed worse compared to (2 s, 2 s). This tells that in case the long segment were to be short, e.g., 2 s, short segments are better to be at least longer than 1 s.

In TABLE~\ref{tab:dino_seg_ratio}, we can observe that having shorter segments decreases the performance with the cosine scoring back-end while increasing the performance with the \ac{PLDA} back-end. The decrease in cosine scoring could happen due to the sample length mismatch during the training and testing, e.g., the average utterance length in VoxCeleb1 \textit{test} is about 8.27 s. However, the mismatch seems to be compensated by the PLDA back-end training that uses the in-domain VoxCeleb1 \textit{dev}, resulting in the better performances than cosine scoring. The reason for the slight gains with \ac{PLDA} when using shorter lengths in the short augmented segments could be adding more diversity in training samples with shorter lengths. In other words, given an utterance, more diverse training segments can be extracted using a shorter segment length.

Finally, we summarize findings in TABLE~\ref{tab:dino_longseg_len} and~\ref{tab:dino_seg_ratio}: 1) Long segments are better to be at least longer than 2 s. 2) In case the long segment were to be short, e.g., 2 s, short segments are better to be at least longer than 1 s. 3) When the long segment is 4 s, cosine scoring performs better in general as the short segment length increases while PLDA performs better as the short segments get shorter.

In later experiments, we use 4 s and 2 s for long and short segments, respectively.

\begin{table}[htbp]
\centering
\caption{\ac{DINO} long segment length study. (long, short) means long segment length and short segment length in the augmented inputs.}
\begin{tabular}{|c|c|c|} 
\hline
(LONG, SHORT) & Cosine scoring & PLDA \\ \hline \hline
(4 s, 2 s) & \textbf{4.83} & \textbf{2.38} \\ 
(3 s, 2 s) & 5.94 & 2.60 \\ 
(2 s, 2 s) & 8.60 & 3.23 \\ \hline
(2 s, 1 s) & 10.56 & 3.42 \\
\hline
\end{tabular}
\label{tab:dino_longseg_len}
\end{table}
\vspace{-0.1in}

\begin{table}[htbp]
\centering
\caption{\ac{DINO} input long-short segment ratio study.}
\begin{tabular}{|c|c|c|} 
\hline
(LONG, SHORT) & Cosine scoring & PLDA \\ \hline \hline
(4 s, 4 s) & 4.66  & 2.49 \\ 
(4 s, 3 s) & \textbf{4.58} & 2.43  \\ 
(4 s, 2 s) & 4.83 & 2.38 \\
(4 s, 1 s) & 6.00 & \textbf{2.25} \\
\hline
\end{tabular}
\label{tab:dino_seg_ratio}
\end{table}

\subsection{Data augmentation}
In this experiment, we checked the importance of data augmentation following the policy in Section \ref{subsec:model_pretraining} during the DINO training. The comparison result between with and without augmentation is shown in Table~\ref{tab:dino_dataaug} based on \ac{EER}(\%), where we found that the noise augmentation led to significant performance improvement. The effectiveness of the noise augmentation was observed in the past for both supervised~\cite{snyder2018xvector} and self-supervised speaker embedding training~\cite{xia2021moco_spkemb}.

The possible reason for the improved performance with augmentation is that adding noise to the segments from the same utterance encourages the model to focus more on information unaffected after noise addition, e.g., speaker information. In contrast, the model might learn channel or noise information when noise augmentation is not applied.

\begin{table}[htbp]
\centering
\caption{Noise augmentation during \ac{DINO} training}
\begin{tabular}{|c|c|c|} 
\hline
 & Cosine scoring & PLDA \\ \hline \hline
NO augmentation & 23.27  & 7.81 \\ 
Noise augmentation & \textbf{4.83} & \textbf{2.38}  \\ 
\hline
\end{tabular}
\label{tab:dino_dataaug}
\end{table}

\subsection{Contrastive VS non-contrastive learning}
\begin{table}[htbp]
\centering
\caption{Comparison of \ac{DINO} to three contrastive learning methods in \ac{SV} on VoxCeleb1 \textit{test}.}
\begin{tabular}{|c|c|}
\hline
Method & EER(\%) \\ \hline \hline
\ac{DINO} & \textbf{4.83} \\ \hline
AP+AAT~\cite{huh2020augmentation} & 8.65 \\
MoCo (ProtoNCE)~\cite{xia2021moco_spkemb} & 8.23 \\
CSSL w/ $L_{ap}+L_{ch(mse)}$~\cite{zhang2021contrast_spkid} & 8.28 \\
\hline
\end{tabular}
\label{tab:cmpr_ssls_sv}
\end{table}
First, we compared DINO as a non-contrastive embedding learning method to three contrastive learning methods in the previous works. The \ac{SV} systems using the learned embeddings and cosine scoring were compared in Table~\ref{tab:cmpr_ssls_sv}. Embedding learning in all the models used the VoxCeleb2 \textit{dev} data with data augmentation but without the speaker labels. The systems were evaluated on VoxCeleb1 \textit{test}. As shown in the table, DINO outperformed all the previous systems based on the contrastive embedding learning.

\subsection{Frozen}
\begin{table*}
\caption{Overview of fine-tuning experiments for \ac{SV}. \textit{FT2}\textsubscript{\textrm{AAM}} means that the \ac{AAM} loss was used in the fine-tuning while \textit{FT2} uses a cross entropy loss. x-vector (-) means that the pre-training of the x-vector system did not use VoxCeleb2 \textit{dev}}
\centering
\begin{tabular}{|c|c|c|c|c|c|c|}
\hline
   & Pretraining(-Finetuning)   & labeled data (pre-training, fine-tuning)      & back-end & labeled data (backend) & EER (\%) & MinDCF(p=0.01) \\ \hline \hline
1  & DINO           & none, none                   & CS    & none    & 4.83     & 0.463  \\
2  &                &             & PLDA  & VoxCeleb1 \textit{dev}   & 2.38     & 0.289  \\ \hline
3  & DINO-\textit{FT2}        & none, VoxCeleb1 \textit{dev}            & CS   & none    & 3.41     & 0.373  \\ 
4  &                &             & PLDA  & VoxCeleb1 \textit{dev}   & 2.20     & 0.278  \\ \hline
5  & DINO-$\textit{FT2}_{\textrm{AAM}}$     & none, VoxCeleb1 \textit{dev}            & CS  & none     & 2.51     & 0.241  \\ 
6  &                &             & PLDA  & VoxCeleb1 \textit{dev}   & 2.01     & 0.255  \\ \hline
7  & x-vector (-)      & VoxCeleb1 \textit{dev}, none            & CS  & none     & 3.32     & 0.345   \\ 
8  &                &             & PLDA  & VoxCeleb1 \textit{dev}   & 2.75     & 0.290   \\ \hline
9  & x-vector       & VoxCeleb2 \textit{dev}, none            & CS    & none   & 2.18     & 0.205  \\ 
10  &                &  & PLDA  & VoxCeleb1 \textit{dev}   & 1.87     & 0.211  \\ \hline
11 & x-vector-\textit{FT2}    & VoxCeleb2 \textit{dev}, VoxCeleb1 \textit{dev} & CS  & none     & 2.46     & 0.298  \\ 
12 &                &  & PLDA  & VoxCeleb1 \textit{dev}   & 1.98     & 0.215  \\ \hline
13 & x-vector-$\textit{FT2}_{\textrm{AAM}}$ & VoxCeleb2 \textit{dev}, VoxCeleb1 \textit{dev} & CS    & none   & 2.21     & 0.215  \\ 
14 &                &  & PLDA  & VoxCeleb1 \textit{dev}   & 1.98     & 0.250  \\ \hline
\end{tabular}
\label{tab:SV_transfer_overview}
\end{table*}
In this experiment, we compared embeddings from x-vector and DINO pre-trained models without any fine-tuning for \ac{SV}. 
Row 1, 2 and 7 to 10 in Table~\ref{tab:SV_transfer_overview} show the results. Firstly, the \ac{SV} system with \ac{DINO} without any labels (row 1) shows 4.83 EER (\%). This can be considered a low EER given that no speaker labels  were used to build the \ac{SV} system.

When we replaced the cosine scoring back-end with \ac{PLDA} trained on VoxCeleb1 \textit{dev} (row 2), it gives a large improvement to 2.38 EER (\%).
Compared to the pre-trained x-vector models in row 7 and 8, which uses the same amount of the labeled data, \ac{DINO} with \ac{PLDA} performs better.
When we compare \ac{DINO} followed by \ac{PLDA} to the x-vector models trained on labeled VoxCeleb2 \textit{dev} (row 9 and 10), x-vector outperforms \ac{DINO}. However, \ac{DINO} used only about 1/7 of the labeled data used in the x-vector model. This is a simulated scenario that is practical, where a large amount of data exist without labels while only a small amount of data is available with labels.

VoxCeleb1 \textit{dev} used for \ac{PLDA} also can be used to fine-tune the \ac{DINO} pre-trained model, whose improved results will be explained in the following section.

\subsection{Fine-tuned}
\begin{table}
\caption{DINO fine-tuning experiments for \ac{SV}}
\centering
\begin{tabular}{|c|c|c|c|c|}
\hline
FT     & Loss    & back-end & EER (\%) & MinDCF(p=0.01) \\ \hline \hline
No FT  & -       & CS       & 4.83     & 0.463  \\
       &         & PLDA     & 2.38     & 0.289  \\ \hline
\textit{FT1}-4s & softmax & CS       & 3.57     & 0.358  \\ 
       &         & PLDA     & 2.54     & 0.290  \\
       & AAM     & CS       & 2.72     & 0.281  \\ 
       &         & PLDA     & 2.47     & 0.282  \\ \hline
\textit{FT1}-2s & softmax & CS       & 3.30     & 0.383  \\ 
       &         & PLDA     & 2.37     & 0.267  \\
       & AAM     & CS       & 2.65     & 0.278  \\ 
       &         & PLDA     & 2.48     & 0.301  \\ \hline
\textit{FT2}-2s & softmax & CS       & 3.41     & 0.373  \\ 
       &         & PLDA     & 2.20     & 0.278  \\
       & AAM     & CS       & 2.51     & 0.241  \\ 
       &         & PLDA     & 2.01     & 0.255  \\ \hline
\end{tabular}
\label{tab:SV_FT_DINO}
\end{table}
In Table~\ref{tab:SV_transfer_overview}, the 2nd, 4th and 6th rows  show that using VoxCeleb1 \textit{dev} also for fine-tuning the pre-trained \ac{DINO} model further improves the results. The \ac{AAM} loss was more helpful than the cross entropy loss in fine-tuning. When only VoxCeleb1 \textit{dev} was used as labeled data, the best \ac{DINO}-based system (row 6) outperforms the x-vector system (row 8) by 0.74 in EER(\%)

The x-vector models fine-tuned on VoxCeleb1 \textit{dev} from the x-vector pre-trained on VoxCeleb2 \textit{dev} (row 11 to 14) outperformed the best x-vector model trained on VoxCeleb1 \textit{dev} from scratch (row 8). The best fine-tuned x-vector model (row 12) was 0.03 better than the best fine-tuned DINO model (row 6) in EER(\%). However, the fine-tuned DINO model only used about 1/7 of the labeled data used in the fine-tuned x-vector model, during the whole training process (pre-training + fine-tuning).

Fine-tuning the pre-trained x-vector (row 11-14) actually worsened the performance slightly, compared to using the pre-trained x-vector as it is (row 9 and 10). This might be because the pre-training and fine-tuning data are in-domain where both are used to train speaker classification models while the fine-tuning data has less speakers than the pre-training data. Thus, in the x-vector case, simply using both labeled data together would perform better than dividing them into pre-training and fine-tuning. An experiment that confirms this is not carried out since this is not the focus of this paper.

Table~\ref{tab:SV_FT_DINO} shows the results over different fine-tuning configurations from \ac{DINO}. We observed that when cosine scoring back-ends were used, using \ac{AAM} during fine-tuning improves performances. This is probably because the \ac{AAM} loss tries to reduce intra-class distances while increasing inter-class distances in terms of cosine distance. However, when \ac{PLDA} was used for back-end, the performances only improved in \textit{FT1}-2s with softmax loss, and all \textit{FT2}-2s systems. This shows that when fine-tuning from the \ac{DINO} model, it is important not to directly fine-tune the whole model with a newly added last affine layer, but rather first fine-tune only a few layers after the pooling layer including the newly added layer, and then fine-tune the whole network together. This could prevent the parameters from deviating too much from the pre-trained model's parameters during fine-tuning. Using 2s chunks was better than 4s chunks during fine-tuning for most of the case in \textit{FT1}. Thus, we only experimented \textit{FT2} with 2s chunks.

\subsection{Iterative pseudo-labeling and fine-tuning}
\begin{table*}[htbp]
\centering
\caption{Speaker verification results over 3 different trial lists with progressing/different systems over the three stages. The numbers from~\cite{thienpondt2020idlab} seems rounded to the nearest tenth. Pseudo labels for robust training were generated from ResNet34 (iter3). No ground truth labels were used during the whole time.}
\vspace{0.1in}
\begin{tabular}{|c|c|c|c|c|c|}
\hline
\multirow{2}{*}{Stage}                    & \multirow{2}{*}{Algorithm/Loss} & \multirow{2}{*}{Model}             & \multicolumn{3}{c|}{EER   (\%) with cosine scoring}         \\ \cline{4-6} 
                                          &                                 &                                    & VoxCeleb1\_test\_o & VoxSRC-21 \textit{val} & VoxSRC-21 \textit{test} \\ \hline \hline
\multirow{2}{*}{\shortstack{Initial   model training \\ (self-supervised learning)}} & DINO                        & LResNet34                          & 4.83     & 13.96        &           -    \\ \cline{2-6} 
                                          & \ac{MoCo}                       & ECAPA~\cite{thienpondt2020idlab}                              & 7.3     &       -       &      -        \\ \hline
\multirow{5}{*}{Iterative clustering}     & \multirow{5}{*}{\shortstack{AAM loss \\ (margin=0.3)}}         & ResNet34 (iter1)                   & 2.56     & 8.59         &         -      \\ \cline{3-6} 
                                          &                                 & ResNet34 (iter2)                   & 2.13     & 7.35         &        -       \\ \cline{3-6} 
                                          &                                 & ResNet34 (iter3) & 2.13     & 6.97         &         -      \\ \cline{3-6} 
                                          &                                 & ResNet34 (iter4)                   & 2.14     & 6.88         &        -       \\ \cline{3-6} 
                                          &                                 & ECAPA (iter7)~\cite{thienpondt2020idlab}                      & 2.1     &    -          &        -       \\ \hline
Robust training                           & \multirow{2}{*}{\shortstack{AAM loss\\(margin=0.5)}}                          & \multirow{2}{*}{Res2Net50}                        & 1.89     & 6.50         & 6.88          \\ \cline{4-6}
+ larg-margin fine-tuning &                         &                          & 1.91     & 6.32         & 6.64          \\ \hline
\end{tabular}
\label{tab:all_results}
\vspace{-5mm}
\end{table*}

The experimental results are shown in Table~\ref{tab:all_results} along the iterative process. In the iterative pseudo-labeling and fine-tuning before the robust training stage, the speaker verification performance is saturated around the $3^{rd}$ iteration. This number of iterations until convergence is less than the one reported in~\cite{thienpondt2020idlab}\footnote{This is a rough comparison since validation sets and detailed configurations are slightly different.}, possibly due to starting from a better initial model. Thus, we generated the pseudo labels from the model after the $3^{rd}$ iteration (ResNet34 (iter3)) to use them for the last model training in the robust training stage, which improved further to 1.89 in EER(\%) on VoxCeleb1\_test\_o. This speaker verification system did not use any speaker labels in the development, and the supervised counterpart trained on about 2600 hours of speaker labeled data showed 0.93. Finally, the large-margin fine-tuning did not improve on the VoxCeleb1\_test\_o trial list, while it improved on VoxSRC-21 \textit{val} and \textit{test} trial lists.

\section{Emotion Recognition Results}

\subsection{Frozen}
\begin{figure}[htbp]
\centering
\includegraphics[width=0.45\textwidth]{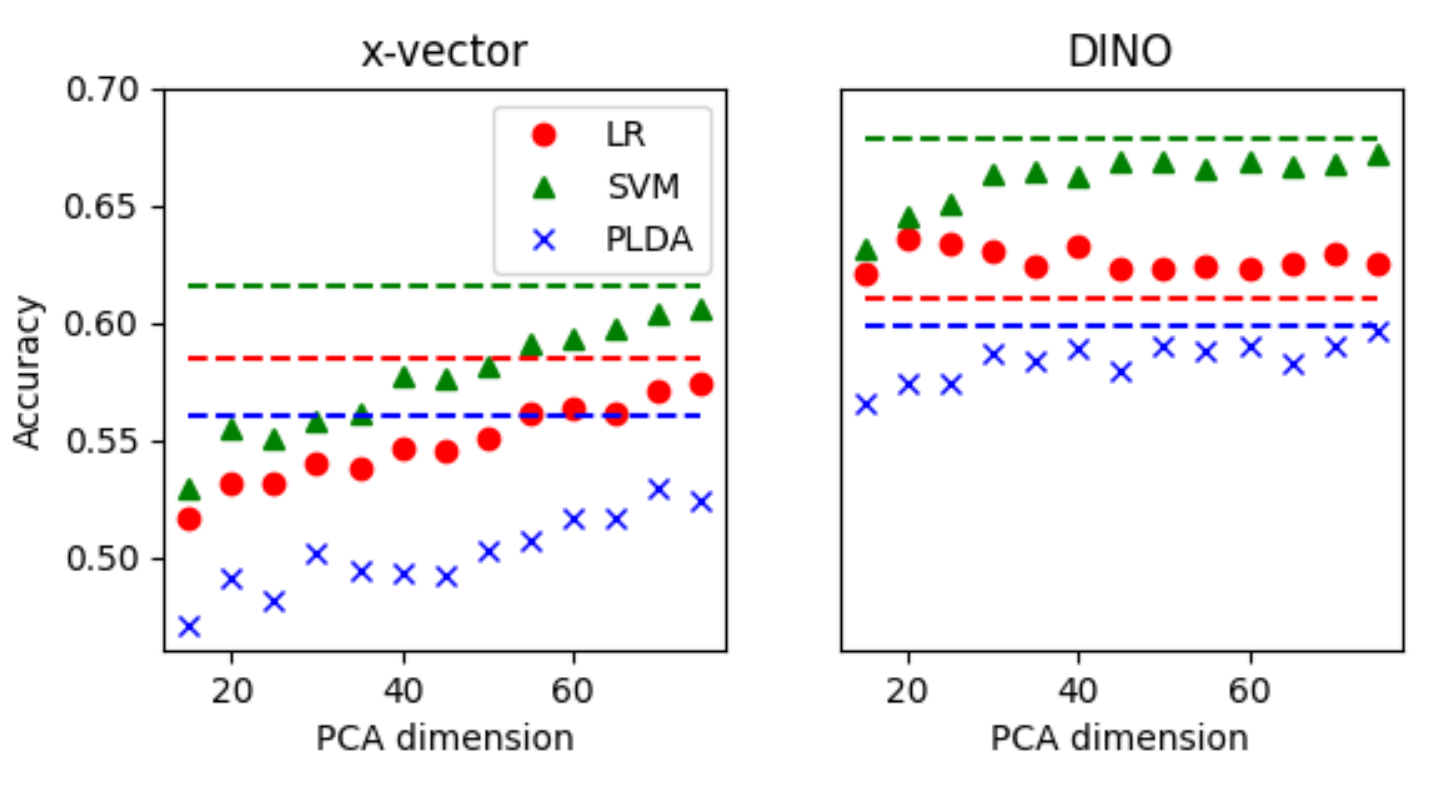}
\caption{Frozen SER. Dotted lines are accuracies when PCA is not used}
\label{fig:Frozen_SER}
\end{figure}
The results for the experiments are shown in Fig. \ref{fig:Frozen_SER}. The \ac{SVM} without \ac{PCA} provides the highest accuracy among three classifiers for both x-vector and DINO frozen models, showing 0.6160 and 0.6788 in accuracy respectively. This implies that the emotion classes are not linearly separable in the embedding spaces, thus making the \ac{SVM} with a RBF kernel work the best.
\ac{PCA} works only when it was used with DINO embeddings followed by the \ac{LR} classifier where it gives improvements at all the \ac{PCA} dimensions.
Comparing the x-vector to \ac{DINO} embeddings, the latter was better in general. This suggests \ac{DINO} retains more information related to emotion.

\subsection{Fine-tuned}
\begin{table}[hbtp]
\caption{Fine-tuning accuracies for \ac{SER} and \ac{AD} detection}
\centering
\begin{tabular}{|c|c|c|}
\hline
Fine-tuning & SER & AD detection \\ \hline \hline
DINO-\textit{FT1}    & 0.6371        & 0.7529           \\ 
DINO-\textit{FT2}    & 0.6656        & 0.7715           \\ 
xvector-\textit{FT2} & 0.6356        & 0.7163           \\ \hline
\end{tabular}
\label{tab:SERnSPD_FT}
\end{table}
%

Firstly, we compared the direct fine-tuning with a newly added layer (\textit{FT1}) with the 2-stage fine-tuning (\textit{FT2}), whose results are shown in Table~\ref{tab:SERnSPD_FT}. In the \ac{DINO} cases, \textit{FT2} outperformed \textit{FT1} by large margin. This trend is the same as we observed in \ac{SV} fine-tuning experiments. When compared to the 2-stage fine-tuning of the x-vector (xvector-\textit{FT2}), DINO-\textit{FT2} performs better, suggesting that the \ac{DINO} model is adapted better to process emotion-related information than x-vector.
\begin{figure}[htbp]
\centering
\includegraphics[width=0.45\textwidth]{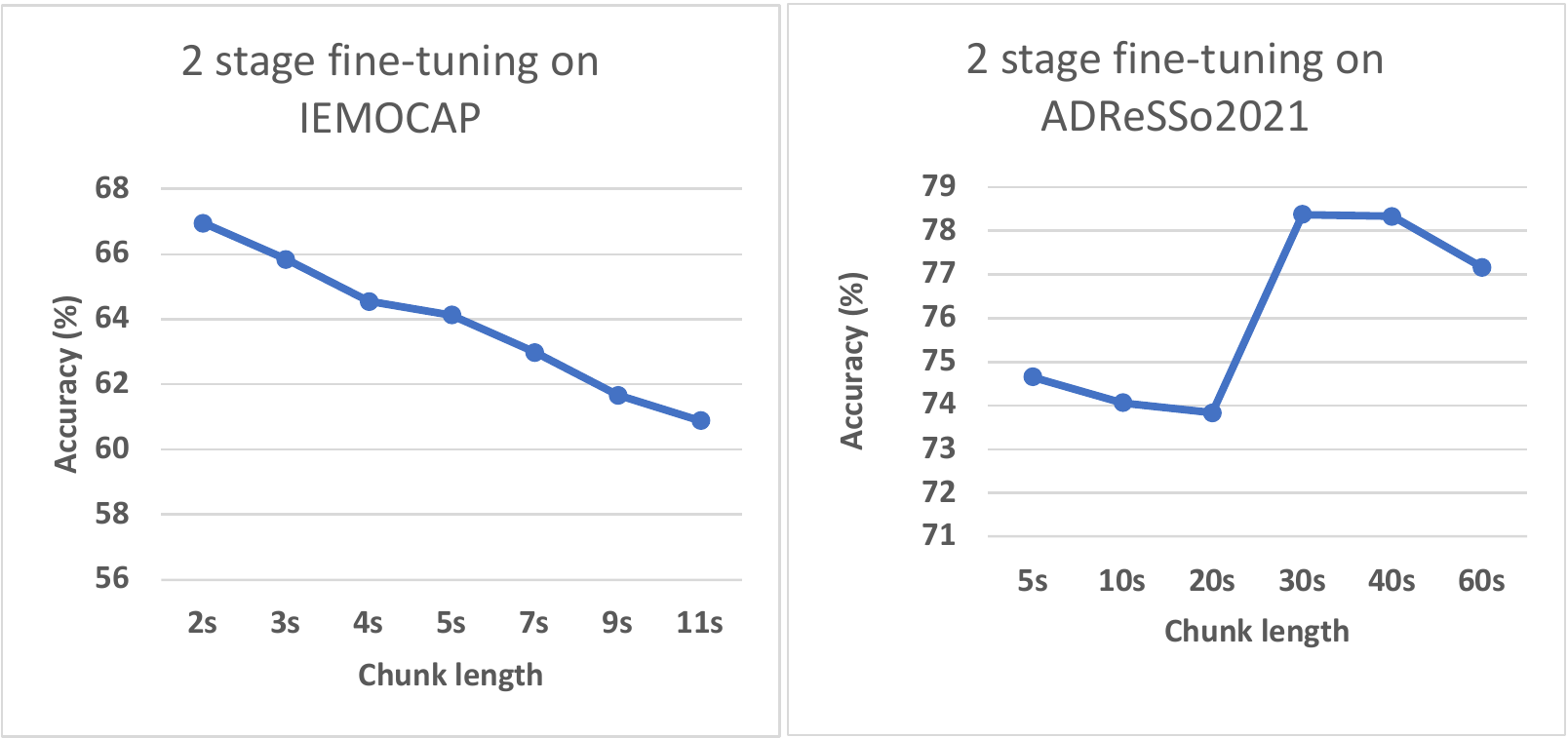}
\caption{Fine-tuning with different chunk length for inputs}
\label{fig:FT_chunklen_zeropad}
\end{figure}
\begin{figure}
\centering
\includegraphics[width=0.45\textwidth]{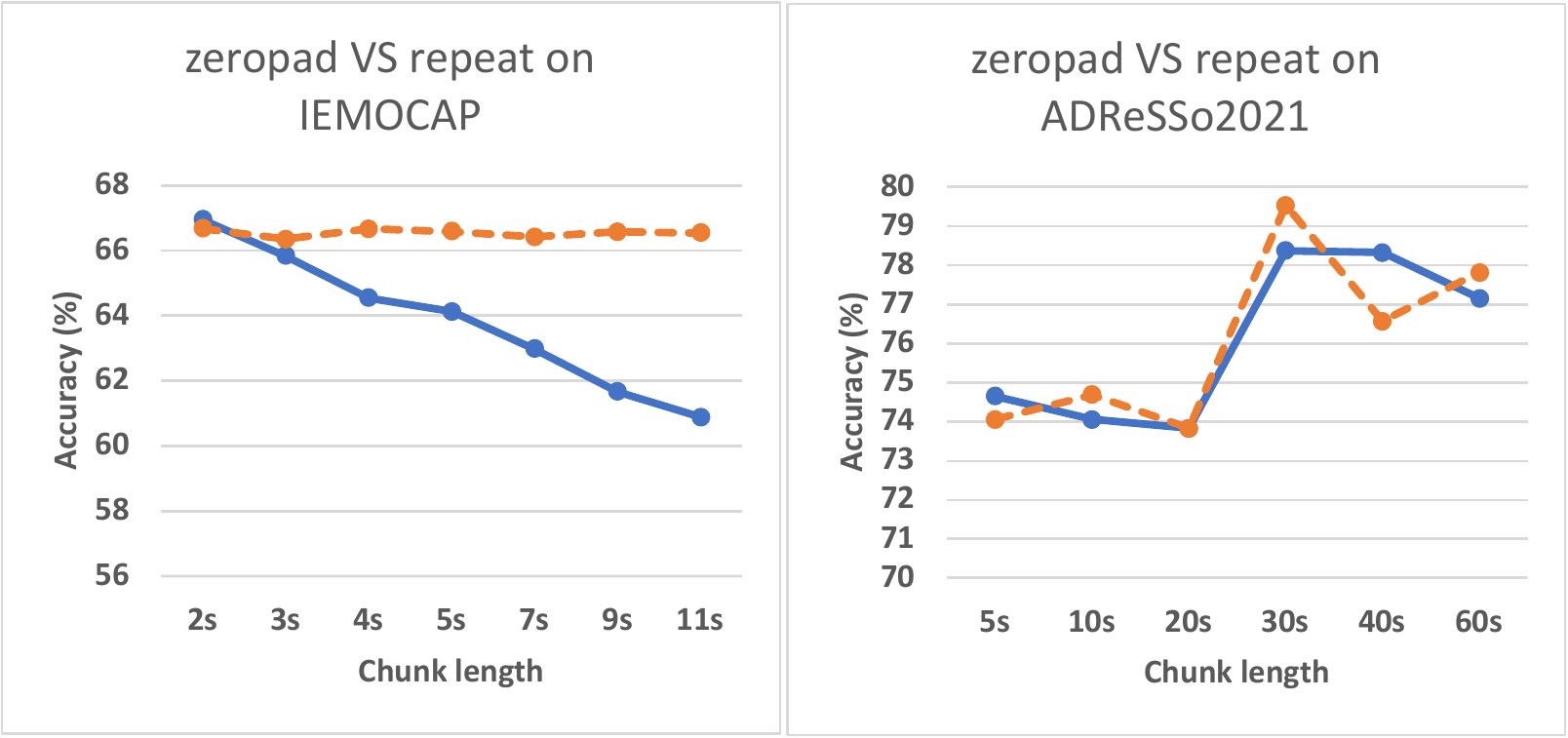}
\caption{2 types of padding experiment in fine-tuning. The orange dotted and the blue solid lines are repeating and zero-padding short utterances, respectively.}
\label{fig:FT_zeropadVSrepeat}
\end{figure}
\begin{figure}
\centering
\includegraphics[width=0.45\textwidth]{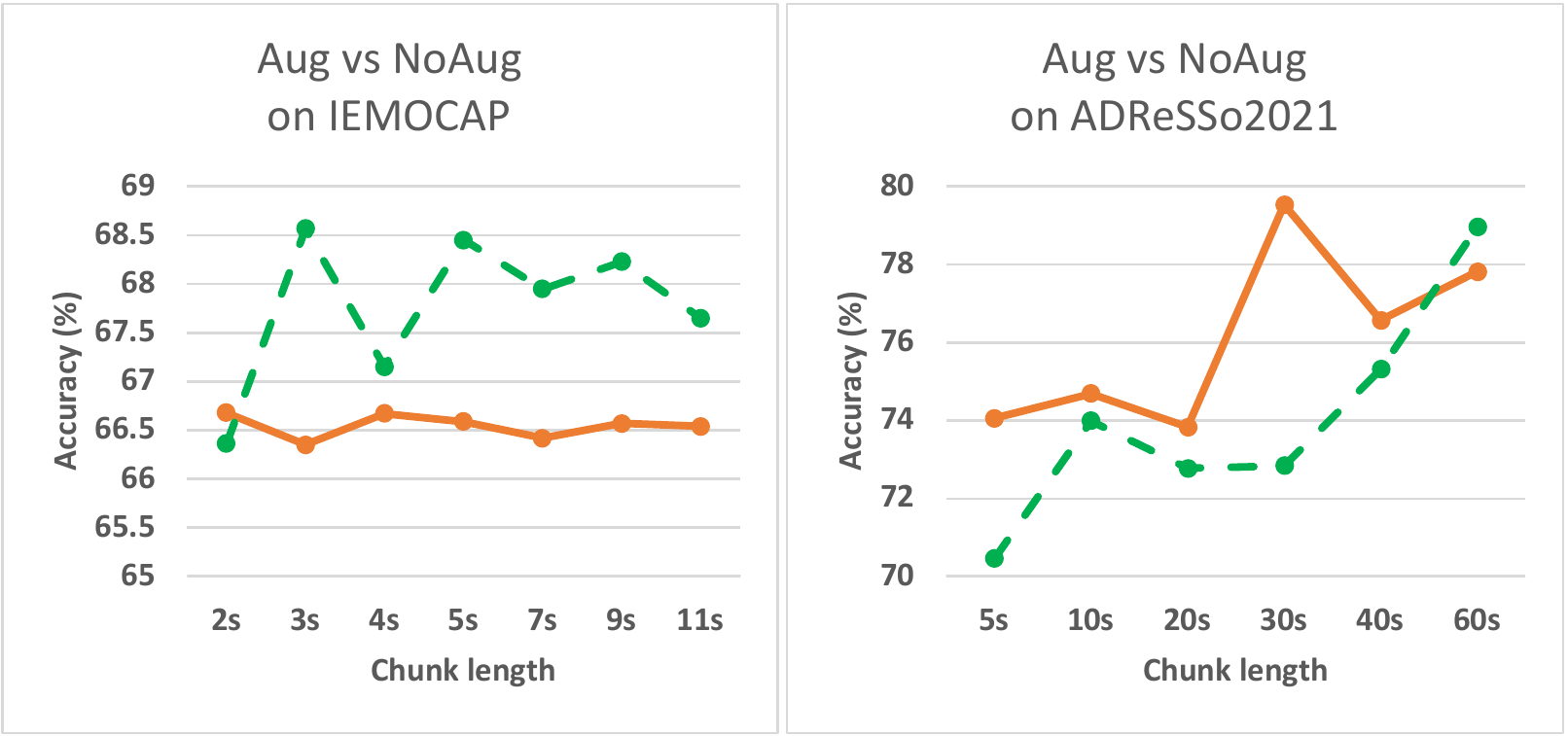}
\caption{Augmentation experiment in fine-tuning. The green dotted and the orange solid lines are augmentation and no augmentation, respectively.}
\label{fig:FT_augVSnoaug}
\end{figure}
%

Next, The left graph in Fig.~\ref{fig:FT_chunklen_zeropad} shows the results from the experiments over different training chunk lengths.
We could observe that the performance consistently degraded as training chunk length increased. In the experiments, the higher the chunk length is, the more zero-padding happens since the number of utterances shorter than the chunk length increases. In other words, padding shorter utterances simply with zeros can harm the training further as it happens more. To fix this zero-padding correctly, masking the affected parts by the padded zeros is required, which entails a more complicated implementation.

As a simpler alternative implementation while not degrading the performance, we instead repeated the short utterances, expecting emotion in the processed utterances to remain the same. The graph comparing the two methods is shown at the left of Fig.~\ref{fig:FT_zeropadVSrepeat}. In all chunk lengths, repeating short utterances outperformed zero-padding them, especially giving more improvement when zero-padding happens more frequently, i.e., as the training chunk length increases. One interesting to note with repeating short utterances (the orange dotted line) is that using short chunk length, e.g., 2s and 3s, does not necessarily work better than using longer chunks. This implies that although the short chunk length has more various samples during training, it needs to be long enough to preserve the emotion information in the chunks.

Finally, we compared how noise data augmentation using the data augmentation policy in Section~\ref{subsec:model_pretraining} affects the fine-tuning. As the results are shown in the left plot in Fig.~\ref{fig:FT_augVSnoaug}, the augmentation helped to improve the results in general.

\section{AD Detection Results}
\subsection{Frozen}
\begin{figure}[htbp]
\centering
\includegraphics[width=0.45\textwidth]{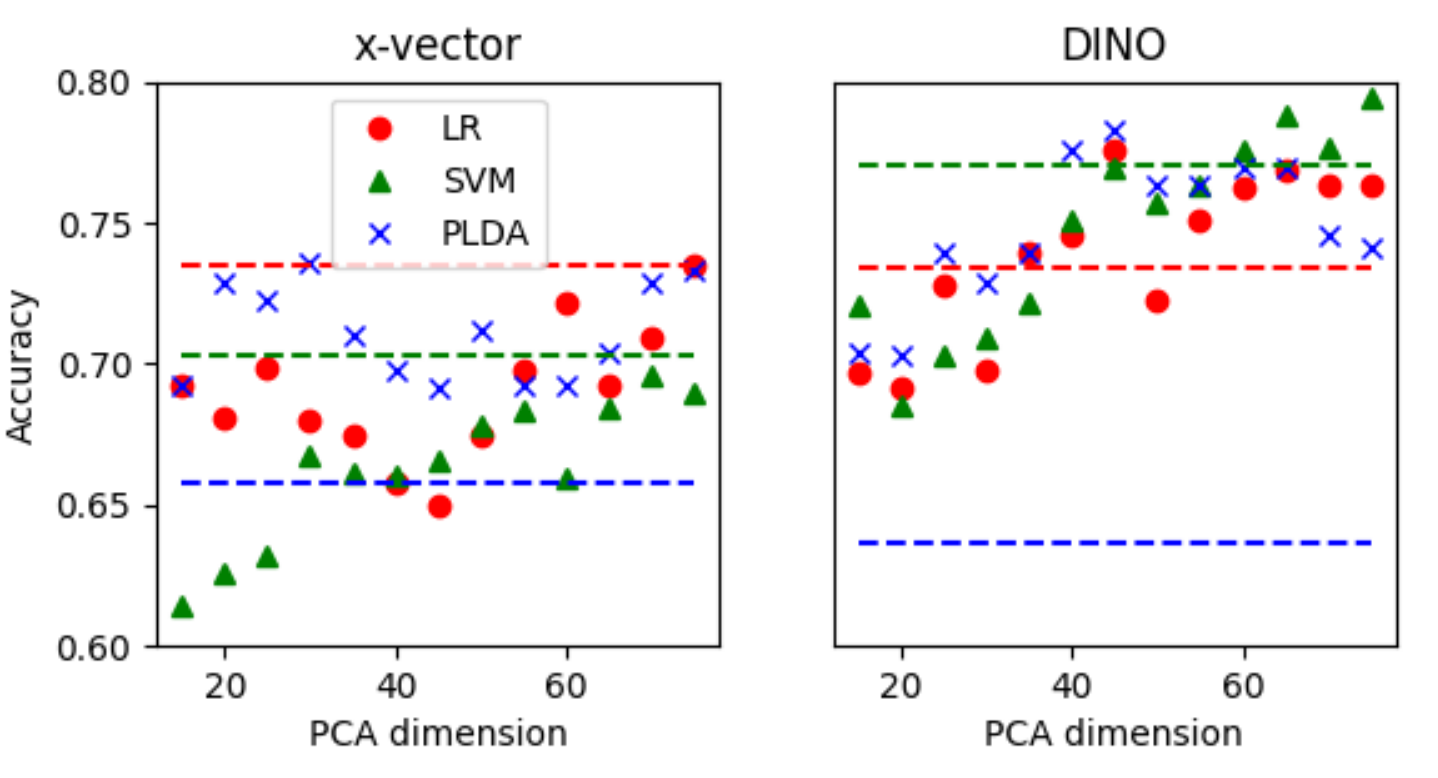}
\caption{Frozen AD detection. Dotted lines are accuracies when PCA is not used}
\label{fig:Frozen_SPD}
\end{figure}
Next, the results on \ac{AD} detection are shown in Fig. \ref{fig:Frozen_SPD}. Here, \ac{PLDA} with \ac{PCA} using the 30 principal components and \ac{SVM} with 75-dimensional \ac{PCA} worked best for x-vector and DINO embeddings with 0.7355 and 0.7943 in accuracy, respectively.
Using \ac{PCA} led to improvement in many cases and to the larger degree compared to when used for \ac{SER}.
\ac{DINO} embeddings work better than x-vector embeddings for \ac{AD} detection in general, implying \ac{DINO} retains more information related to \ac{AD} detection.

\subsection{Fine-tuned}
As shown in the AD detection (ADReSSo2021) column in Table~\ref{tab:SERnSPD_FT}, the 2-stage fine-tuning outperformed fine-tuning at once from the \ac{DINO} model, possibly for the same reason that we explained in the \ac{SER} experiment. Also, fine-tuning the \ac{DINO} model showed a better result compared to fine-tuning the x-vector model.

The results from experiments with different chunk lengths are shown in the right graph of Fig.~\ref{fig:FT_chunklen_zeropad}. Our hypothesis is that there were performance drops for 2 reasons. First, in 5s, 10s, and 20s, the lengths are too short for a model to extract information relative to \ac{AD}, that is related to articulatory rate, speech/pause rate, or presence of hesitations, among others. In 60s, zero-padding happens more due to increase in the number of utterances shorter than 60s, according to the ADReSSo2021 length duration histogram in \ref{fig:utt_dur_hist} . This was observed similarly in the \ac{SER} experiments. 

As it is done in the \ac{SER} experiments, we replaced the zero-padding by repeating the utterances. The results are shown in the right graph in Fig.~\ref{fig:FT_zeropadVSrepeat}. Different from observation in the \ac{SER} experiments, repeating did not help much in general except at several lengths. Our hypothesis is that this observation is more related to that 5s, 10s, and 20s-length chunks are too short for a model to detect \ac{AD}, than the padding option.

Lastly, we compared how noise data augmentation affects the fine-tuning in \ac{AD} detection. As the results are shown in the right graph in Fig.~\ref{fig:FT_augVSnoaug}, the augmentation worsened the performance in general. One possible reason is that ADReSSo2021 is already noisy and adding more noise could be lowering the quality of the recordings.

\section{Conclusion}
In this work, we applied \ac{DINO}, a non-contrastive self-supervised learning method first proposed in computer vision, to speech domain. We compared a model pre-trained with this method to an x-vector model trained in a supervised manner, w.r.t transfer learning to three speech applications: \ac{SV}, \ac{SER}, and \ac{AD} detection. In SV, \ac{DINO} enabled building \ac{SV} systems without speaker labels, and improved x-vector by providing a better initialized model for the training. This implies that \ac{DINO} effectively learns speaker information without labels. In \ac{SER} and \ac{AD} detection, \ac{DINO} outperformed x-vector in most of the experiments, suggesting that \ac{DINO} learns embedding that includes more information related to emotion and \ac{AD} than x-vector. From these findings, we conclude that the \ac{DINO} embedding is more general and transferable utterance-level embedding than x-vector for speech applications that require the corresponding utterance-level information.

In the SV experiments, when the \ac{DINO} embedding without fine-tuning and x-vector were compared, \ac{DINO} outperformed x-vector given the same amount of labeled data during their training. When the initial \ac{DINO} model was fine-tuned based on the iterative pseudo-labeling and fine-tuning process, it improved further to 1.89 in EER(\%). This is noteworthy, considering that the whole training pipeline did not use any speaker labels. The initial \ac{DINO} model also can be fine-tuned with a small amount of labeled data using the x-vector training scheme, which can be applied in frequent real-world scenarios where only a small portion of the large amount of data is labeled. The results showed that the x-vector model trained from the \ac{DINO} initialized weights outperformed the x-vector model trained from scratch given the same amount of labeled data. These findings suggest that \ac{DINO} learns speaker information effectively without speaker ID labels.

In the experiments that transfer \ac{DINO} or x-vector model to \ac{SER} and \ac{AD} detection tasks, the best performing models over all experiments were DINO-based models, suggesting \ac{DINO} learns more transferable representation to those speech applications. When using pre-trained models as feature extractors, we found that \ac{PCA} improves the performance in the \ac{AD} detection task.
In other experiments that fine-tune the whole pre-trained models with a newly added affine layer per speech task, we found that fine-tuning affine layers first and then the whole network works (\textit{FT2}) better than fine-tuning the whole network in one step. We think that when the pre-trained model is fine-tuned together with the newly added affine layer that is randomly initialized, large gradients from the new layer flow back to the whole networks to possibly change the parameters much than needed. As a result, this might not utilize the knowledge in the pre-trained model learned from a lot of data, and \textit{FT2} could resolve the problem.

Also, we studied several aspects to analyze how each aspect affects the fine-tuning. First, we checked the chunk length in fine-tuning a model using a specified-length segments extracted at a random position of a given utterance. We found that although shorter chunk length makes training samples more diverse, it did not necessarily improve the performance. In the \ac{AD} detection task, performance started to improve from 30-second of chunk length, suggesting that at least 30-second of speech segment is required to capture the relevant information. Along the experiments, we found that simply zero-padding utterances shorter than specified chunk length could degrade the performance as it happens more, thus replacing them by repeating utterances for the shorter ones. In noise data augmentation experiments, the augmentation only improved the performance in \ac{SER} task, but not in \ac{AD} detection.

\section{Discussion}
Although we applied noise data augmentation following the same policy to all speech applications, it would be interesting to analyze which types of data augmentation is helpful depending on speech application, among augmentation methods more diverse than ones introduced in this paper such as speed perturbation, VTLP, and specaugmentation. Another possible future work is to see how a larger model, longer epoch, or more data in \ac{DINO} pre-training is related to the performance of \ac{DINO} in fine-tuning. As another future work, we could keep the \ac{DINO} training design for fine-tuning. In \ac{SER}, for example, we can concatenate all the utterances in an emotion class to get short and long segments from the long concatenated sequence. Then, \ac{DINO} can maximize the similarity between feature distributions of the segments because they are all from the same emotion. \ac{DINO} can be also evaluated on other tasks that require utterance-level information such as age estimation and language/accent classification. Finally, we could multi-task learn a speaker ID + emotion (or + AD) classifier to reduce data scarcity problem, which uses a large amount of data with speaker labels in addition to a small amount of the labeled data for a target application.

\vfill
\bibliographystyle{myIEEEbib}
\bibliography{refs}

\end{document}